# Solving Schrödinger Equation with Scattering Matrices. Bound states of Lennard-Jones Potential

Carlos Ramírez[1]*, Fernanda H. González[1], and César G. Galván[2]

[1] *Departamento de Física, Facultad de Ciencias, Universidad Nacional Autónoma de México, Apartado Postal 70542, 04510 Ciudad de México, México*
[2]*Facultad de Ciencias, Universidad Autónoma de San Luis Potosí, Apartado Postal 78000 San Luis Potosí, México*

This paper presents an accurate highly efficient method for solving the bound states in the one-dimensional Schrödinger equation with an arbitrary potential. We show that the bound state energies of a general potential well can be obtained from the scattering matrices of two associated scattering potentials. Such scattering matrices can be determined with high efficiency and accuracy, leading us to a new method to find the bound state energies. Moreover, it allows us to find the associated wavefunctions, their norm, and expected values. The method is validated by comparing solutions of the harmonic oscillator and the hydrogen atom with their analytical counterparts. The energies and eigenfunctions of Lennard-Jones potential are also computed and compared to others reported in the literature. This method is highly parallelizable and produces results that reach machine precision with low computational effort. A parallel implementation of this method written in FORTRAN is included in the Supplemental material to solve eigenstates of the Lennard-Jones potential.

## 1. Introduction

The physics of the one-dimensional systems has attracted the attention of the scientist in the last years because of the great number of applications in electronics and optoelectronics. One-dimensional devices, such as FETs, diodes, photo-detectors, LEDs, etc, are of great importance in the development of technology and renewable energy.[1] Solutions of the one-dimensional Schrödinger equation (1DSE) are also relevant to describe vibrational states in diatomic molecules within the Born-Oppenheimer approximation, where interatomic effective potentials are frequently modeled by Lennard-Jones, Morse, Morse/Long-range and other potentials.[2-5] On the other hand, solution of the radial equation of Yukawa potential is useful to describe the interaction between a pair of nucleons, particles immersed in plasma, colloidal particles in electrolytes, or charged particles in a sea of conduction bands.[6,7] Most of these potentials have not analytical solution.

The bound states of the potential well have been studied by means of Airy functions[8] and Monte Carlo methods[9] leading to analytical solutions of specific potentials. On the other hand, the transfer matrix has been used to locally solve periodic potentials for different kind of waves.[10,11] Bound states can be solved in terms of a tunneling problem, by adding a barrier on both sides of the quantum well.[12] Other methods for solving arbitrary potentials use the transfer matrix where the potential well is divided into flat barriers[13,14] which leads us to better results than those given by the WKB approximation and variational calculations.[15] Such methods are also used to find recursive solutions of the 1DSE.[16] On the other hand, the scattering matrix (S-Matrix) method has been used to study the time dependent scattering of wave packets to obtain the bound states in 1DSE. These include the step and two delta-function potentials,[17] the square well potential[18] and those that use graphic constructions[19] on the bases of the Pitkanen method.[20] Approximate analytic expression of the ground state energy and its wavefunction has been developed by converting the 1DSE into the non-linear Ricatti equation.[21] A method for solving the 1DSE eigenvalue problem for the anharmonic oscillator using a Liouville transformation has also been proposed, finding uniform convergence in a problem where perturbation theory fails. [22]

In a recent work,[23] we propose a method to find the scattering states of an arbitrary potential in one dimension. In this method, the scattering region is divided into $N$ smaller slices. In each slice the potential is approximated by its truncated Taylor series. This method is highly accurate, numerically stable and requires low computational effort. In this work, we propose a new algorithm to exploit these advantages during calculations of S-matrices and find the bound states of an arbitrary potential.

The article is organized in the following form. Section 2 describes the algorithm employed to find the bound states of a general potential well in terms of the S-matrix of two associated potentials. In §3, we validate this method by comparing results of the quantum harmonic oscillator and the hydrogen atom. Finally, in §4 we use this method to obtain the energies and normalized eigenfunctions of the Lennard-Jones potential. The Supplemental material contains a FORTRAN implementation of this method to solve the Lennard-Jones potential.[28]

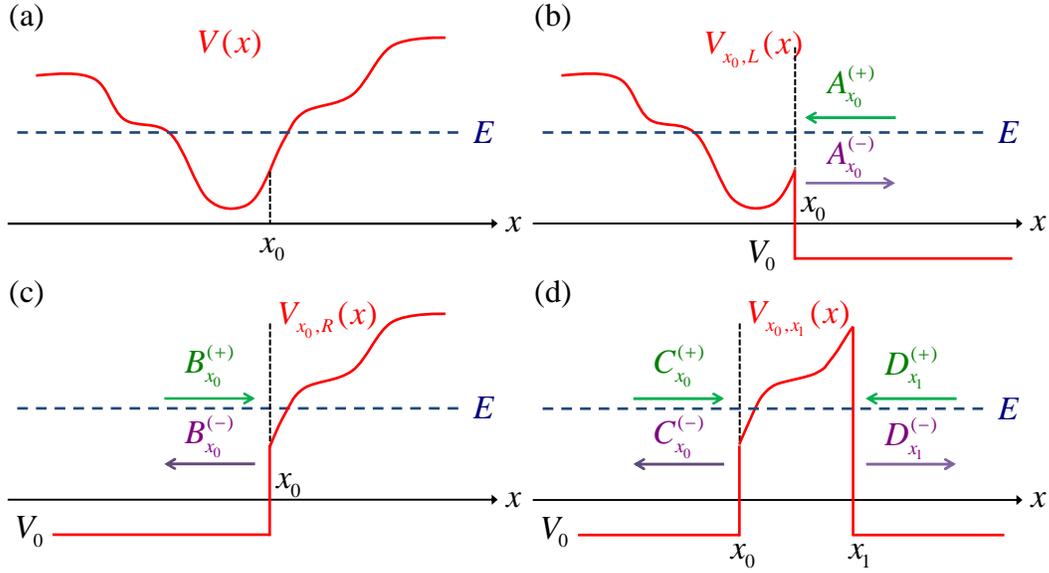

Fig. 1. (a) General one-dimensional potential well. (b) Left associated potential. (c) Right associated potential. (d) Slice associated potential.

## 2. The method

Let us start with a general one-dimensional potential well $V(x)$, as schematically shown in Fig. 1(a), with bound state energies $E_n$ and wavefunctions $\psi_n(x)$. Let us define the following associated potentials

$$V_{x_0,L}(x) \equiv \begin{cases} V(x) & \text{if } x < x_0 \\ V_0 & \text{if } x \geq x_0 \end{cases}, \quad (1)$$

$$V_{x_0,R}(x) \equiv \begin{cases} V_0 & \text{if } x < x_0 \\ V(x) & \text{if } x \geq x_0 \end{cases}, \quad (2)$$

and

$$V_{x_0,x_1}(x) \equiv \begin{cases} V_0 & \text{if } x < x_0 \text{ or } x > x_1 \\ V(x) & \text{if } x_0 \leq x \leq x_1 \end{cases}, \quad (3)$$

with $V_0 < E$, as respectively illustrated in Fig. 1(b), 1(c) and 1(d).

Wavefunctions of potentials (1) to (3) are scattering states that can be written as

$$\psi_{x_0,L}(x) \equiv \begin{cases} \varphi(x) & \text{if } x \leq x_0 \\ A_{x_0}^{(+)} e^{-ik(x-x_0)} + A_{x_0}^{(-)} e^{ik(x-x_0)} & \text{if } x \geq x_0 \end{cases}, \quad (4)$$

$$\psi_{x_0,R}(x) \equiv \begin{cases} B_{x_0}^{(+)} e^{ik(x-x_0)} + B_{x_0}^{(-)} e^{-ik(x-x_0)} & \text{if } x \leq x_0 \\ \varphi(x) & \text{if } x \geq x_0 \end{cases}, \quad (5)$$

and

$$\psi_{x_0,x_1}(x) \equiv \begin{cases} C_{x_0}^{(+)} e^{ik(x-x_0)} + C_{x_0}^{(-)} e^{-ik(x-x_0)} & \text{if } x \leq x_0 \\ \varphi_{x_0,x_1}(x) & \text{if } x_0 \leq x \leq x_1 \\ D_{x_1}^{(+)} e^{-ik(x-x_1)} + D_{x_1}^{(-)} e^{ik(x-x_1)} & \text{if } x \geq x_1 \end{cases}, \quad (6)$$

respectively, where $k \equiv \sqrt{2m(E-V_0)}/\hbar$. Amplitude coefficients $\left(A_{x_0}^{(\pm)}, B_{x_0}^{(\pm)}, C_{x_0}^{(\pm)} \text{ and } D_{x_1}^{(\pm)}\right)$ are related by their corresponding S-matrices as follows

$$A_{x_0}^{(-)} = S^{x_0,L} A_{x_0}^{(+)}, \quad (7)$$

$$B_{x_0}^{(-)} = S^{x_0,R} B_{x_0}^{(+)} \quad (8)$$

and

$$\begin{pmatrix} C_{x_0}^{(-)} \\ D_{x_1}^{(-)} \end{pmatrix} = \mathbf{S}^{x_0,x_1} \begin{pmatrix} C_{x_0}^{(+)} \\ D_{x_1}^{(+)} \end{pmatrix} = \begin{pmatrix} S_{11}^{x_0,x_1} & S_{12}^{x_0,x_1} \\ S_{21}^{x_0,x_1} & S_{22}^{x_0,x_1} \end{pmatrix} \begin{pmatrix} C_{x_0}^{(+)} \\ D_{x_1}^{(+)} \end{pmatrix}. \quad (9)$$

Hereinafter, definitions of potentials, wavefunctions of scattering states, amplitude coefficients, and S-matrices will follow the notation given in eqs. (1) to (9).

Notice that by setting $E = E_n$ and

$$A_{x_0}^{(\pm)} = B_{x_0}^{(\mp)} = \frac{ik\psi_n(x_0) \mp \psi_n'(x_0)}{2ik}, \quad (10)$$

continuity of wavefunction and its derivative imply $\varphi(x) = \psi_n(x)$ in eqs. (4) and (5). In other words, scattering states of potentials (1) and (2) can be employed to describe bound states of potential $V(x)$.

Since wavefunctions $\psi_n(x)$ are bound states, they must have an evanescent behavior if $x \to \pm\infty$. In consequence, they are represented by total reflection states in eqs. (4) and (5), *i.e.*, $S^{x_0,L}$ and $S^{x_0,R}$ are phase factors. They can be independently calculated as functions of energy $E$. Consequently, there is a bound state of potential $V(x)$ at energy $E$ if the first equality in eq. (10) is satisfied, which imply

$$S^{x_0,L} S^{x_0,R} = 1. \quad (11)$$

In general, $S^{x_0,L} S^{x_0,R}$ is a unit complex number. Then, we can calculate eigenenergies of potential $V(x)$ by finding energies for which $\text{Im}(S^{x_0,L} S^{x_0,R}) = 0$ and discarding those with

$\text{Re}\left(S^{x_0,L}S^{x_0,R}\right) = -1$. This can be done very efficiently by using secant or regula falsi methods. Next, we discuss how to calculate $S^{x_0,L}$, $S^{x_0,R}$ and wavefunctions $\psi_n(x)$ for arbitrary potentials.

*2.1 Phase factors*

Let us divide potential $V_{x_0,R}(x)$ in two independent potentials, $V_{x_0,x_1}(x)$ and $V_{x_1,R}(x)$. By using the method in Ref. 23, we can calculate the S-matrix $\mathbf{S}^{x_0,x_1}$ associated to potential $V_{x_0,x_1}(x)$. If $S^{x_1,R}$ is known, then by taking $D_{x_1}^{(\pm)} = B_{x_1}^{(\mp)}$, we obtain

$$S^{x_0,R} = \left( S_{11}^{x_0,x_1} + \frac{S_{12}^{x_0,x_1} S_{21}^{x_0,x_1} S^{x_1,R}}{1 - S^{x_1,R} S_{22}^{x_0,x_1}} \right). \quad (12)$$

There are several potentials for which $S^{x_1,R}$ can be exactly calculated, (a) the infinite potential barrier, (b) the step potential and (c) the semi-infinite periodic potential, illustrated in Fig. 2. The first two can be determined analytically to be respectively

$$S_{\text{barrier}}^{x_1,R} = -1, \quad (13)$$

and

$$S_{\text{step}}^{x_1,R} = \frac{ik + \kappa}{ik - \kappa}, \quad (14)$$

where $\kappa \equiv \sqrt{2m(V_1 - E)}/\hbar$ and $E < V_1$. On the other hand, the S-matrix for the semi-infinite periodic potential can be calculated from the S-matrix of its unit cell.[24]

If $S^{x_1,R}$ is not known, notice that for $x_1$ big enough, the evanescent behavior of the wavefunction imply that $\left|S_{12}^{x_0,x_1}\right| < \varepsilon \ll 1$, and then

$$C_{x_0}^{(-)} = S_{11}^{x_0,x_1} C_{x_0}^{(+)} + S_{12}^{x_0,x_1} D_{x_1}^{(+)} \approx S_{11}^{x_0,x_1} C_{x_0}^{(+)}, \quad (15)$$

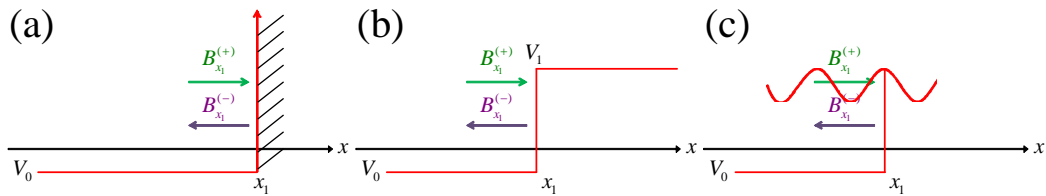

Fig. 2. Potentials for which the phase change can be easily calculated: (a) an infinite potential barrier, (b) a potential step and (c) a semi-infinite periodic potential.

i.e., $S^{x_1,R} \approx S_{11}^{x_0,x_1}$. Since the method in Ref. 23 allow us to iteratively increase the value of $x_1$, we can always make $\varepsilon$ as small as desired to find an excellent approximation of $S^{x_1,R}$.

An analogous procedure may be followed to determine $S^{x_0,L}$ of potential $V_{x_0,L}(x)$.

*2.2 Wavefunctions*

Given an eigenenergy $E_n$ of potential $V(x)$ and the phase factor $S^{x_0,R}$, let us consider potential $V_{x_0,\tilde{x}-\Delta}(x)$ with $\Delta \ll 1$ and $\tilde{x} > x_0$.

Scattering wavefunctions of the 1DSE with potential $V_{x_0,\tilde{x}-\Delta}(x)$ can be written for $-\Delta \le x - \tilde{x} \le \Delta$ as[23)]

$$\psi_n(x) = \alpha_+ \psi_+(x) + \alpha_- \psi_-(x). \quad (16)$$

Functions $\psi_\pm(x)$ can be expressed as

$$\psi_\pm(x) = e^{\pm i\sqrt{v_0}(x-\tilde{x})} \sum_{\lambda=0}^{\Lambda} \varphi_\lambda^{(\pm)} (x-\tilde{x})^\lambda, \quad (17)$$

where $\varphi_0^{(\pm)} = 1$, $\varphi_1^{(\pm)} = \varphi_2^{(\pm)} = 0$, and other coefficients $\varphi_\lambda^{(\pm)}$ are recursively obtained as

$$\varphi_n^{(\pm)} = \mp \frac{2i\sqrt{v_0}}{n} \varphi_{n-1}^{(\pm)} - \frac{1}{n(n-1)} \left( v_{n-2} + \sum_{\lambda=3}^{n-3} \varphi_\lambda^{(\pm)} v_{n-\lambda-2} \right). \quad (18)$$

In eq. (18), $v_n$ are coefficients of the Taylor's expansion

$$\frac{2m[E - V(x)]}{\hbar^2} \equiv \sum_{\mu=0}^{M} v_\lambda (x-\tilde{x})^\lambda, \quad (19)$$

which may be exactly calculated by using automatic differentiation.[25)] Even though $\Lambda = M = \infty$, because $\Delta \ll 1$, finite values of $\Lambda$ and $M$ are enough to reach machine precision during evaluation of wavefunctions.

On the other hand, by using eq. (8) we have

$$C_{x_0}^{(+)} = A \quad \text{and} \quad C_{x_0}^{(-)} = S^{x_0,R} A, \quad (20)$$

where $A$ is a constant. From eq. (20) and the S-matrix $\left(\mathbf{S}^{x_0,\tilde{x}-\Delta}\right)$ of potential $V_{x_0,\tilde{x}-\Delta}(x)$, we obtain

$$D_{\tilde{x}-\Delta}^{(+)} = \frac{S^{x_0,R} - S_{11}^{x_0,\tilde{x}-\Delta}}{S_{12}^{x_0,\tilde{x}-\Delta}} A, \quad (21)$$

and

$$D_{\tilde{x}-\Delta}^{(-)} = S_{21}^{x_0,\tilde{x}-\Delta} A + S_{22}^{x_0,\tilde{x}-\Delta} D_{\tilde{x}-\Delta}^{(+)}. \quad (22)$$

Moreover, by using continuity of wavefunction and its derivative at $x = \tilde{x} - \Delta$, we have

$$\begin{pmatrix} \psi_+(\tilde{x}-\Delta) & \psi_-(\tilde{x}-\Delta) \\ \psi'_+(\tilde{x}-\Delta) & \psi'_-(\tilde{x}-\Delta) \end{pmatrix} \begin{pmatrix} \alpha_+ \\ \alpha_- \end{pmatrix} = \begin{pmatrix} D^{(-)}_{\tilde{x}-\Delta} + D^{(+)}_{\tilde{x}-\Delta} \\ ik\left(D^{(-)}_{\tilde{x}-\Delta} - D^{(+)}_{\tilde{x}-\Delta}\right) \end{pmatrix}. \tag{23}$$

Hence, solving eq. (23), by using eqs. (17), (21) and (22), and substituting into eq. (16) allow us to find an expansion of the wavefunction $\psi_n(x)$. In practice, this expansion is given in terms of a product of exponential and a truncated series of degree $\Lambda$, reaching machine precision if $\Lambda$ is large enough. By varying the value of $\tilde{x}$ we obtain the wavefunction, $\psi_n(x)$, as a piecewise function. It is important to mention that this method to find wavefunctions is valid if

$$\left| S^{x_0,R} - S^{x_0,\tilde{x}-\Delta}_{11} \right| > \tilde{\varepsilon} \tag{24}$$

with $\tilde{\varepsilon}$ greater than machine precision. Otherwise, eq. (21) may lead to numerical errors.

Piecewise expressions of $\psi_n(x)$ also allow us to determine its norm easily. By substituting exponential function in eq. (17) with its Taylor expansion, $\psi_\pm(x)$ can be written as a truncated series of degree $\Lambda$. Consequently, $\psi_n(x)$ and $|\psi_n(x)|^2$ may also be written as truncated series of degree $\Lambda$, which are easy to integrate. Finally, the norm can be obtained by doing a piecewise integration of $|\psi_n(x)|^2$. An analogous procedure may be followed to calculate expected values.

This procedure allows us to find the wavefunction and its norm for $x > x_0$. Analogously, we can obtain the wavefunction and its norm for $x < x_0$.

*2.3 Symmetric potentials*

For $V(x) = V(-x)$ and $x_0 = 0$, we have $S^{0,R} = S^{0,L}$. Then eq. (11) imply that $\left(S^{0,R}\right)^2 = 1$. Consequently, in this case there is a bound state when

$$\text{Im}\left(S^{0,R}\right) = 0. \tag{25}$$

It is worth mentioning that in such zeroes, $\text{Re}\left(S^{0,R}\right)$ is equal to $1$ or $-1$, which correspond to bound states with even or odd wavefunctions, respectively.

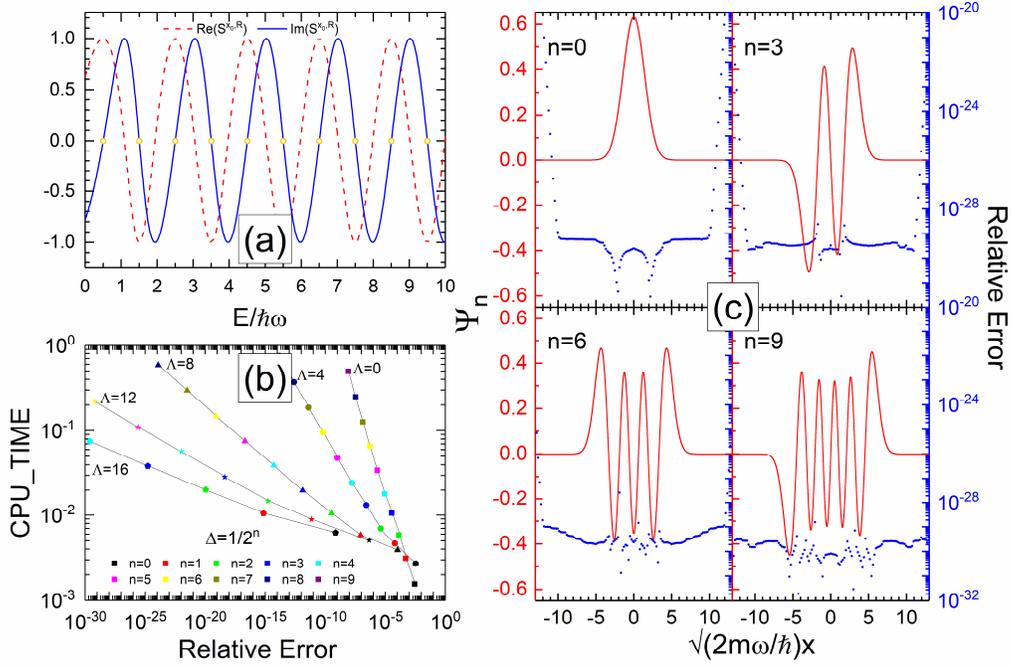

Fig. 3. (a) Calculated imaginary (blue solid line) and real (red dashed line) parts of $S^{0,R}$ as functions of energy $E$ for the quantum harmonic oscillator. (b) Computational time in arbitrary units versus relative error during calculations of the 9th excited state for different values of $\Delta$ and $\Lambda$. (c) Calculated wavefunctions (red solid line) and its relative error (blue dots) for $n = 0, 3, 6$ and $9$.

## 3. Method validation

The one-dimensional harmonic oscillator and the hydrogen atom are two well-known potentials with analytical energies and eigenfunctions.[26,27] We use these potentials to obtain energies and eigenfunctions by using the method in §2 and compare them with their analytical counterparts.

### 3.1 Harmonic Oscillator

The harmonic oscillator, with Hamiltonian

$$\hat{H} = \frac{\hat{p}^2}{2m} + \frac{1}{2}m\omega^2 \hat{x}^2, \qquad (26)$$

has a symmetric potential. Fig. 3(a) shows imaginary (blue solid line) and real (red dashed line) parts of $S^{0,R}$ as a function of energy $E$. Calculation of $S^{0,R}$ was done as in eq. (15)

with $x_1$ big enough to make $\left|S_{12}^{0,x_1}\right|<\varepsilon$. To this end, the S-matrix $\mathbf{S}^{0,2m\Delta}$ $(m=1,2,\cdots)$ was determined iteratively using the method in Ref. 23. Fig. 3(a) was obtained by taking $\varepsilon=10^{-80}$, $\Delta=0.1$, $\Lambda=10$ and $M=2$. Observe that $\mathrm{Im}(S^{0,R})=0$ around the eigenenergies of the harmonic oscillator $E_n^{\mathrm{exact}}=\hbar\omega(n+\tfrac{1}{2})$, shown as yellow open circles in Fig. 3(a). Also notice that the behavior of $\mathrm{Im}(S^{0,R})$ could be well approximated by straight lines around such zeroes. This fact allows us to find accurate approximations of energies, $E_n$, in a few iterations of secant or regula falsi methods. Fig. 3(b) illustrates the computational time and relative error obtained during calculations of $E_9$,

$$\text{Relative Error} \equiv \left|\frac{E_9 - E_9^{\mathrm{exact}}}{E_9^{\mathrm{exact}}}\right|, \quad (27)$$

by applying the secant method, starting with initial seeds at $9.3\hbar\omega$ and $9.6\hbar\omega$, and taking $\varepsilon=10^{-80}$, $M=2$, and different values of $\Delta$ and $\Lambda$. These calculations were done using quadruple precision arithmetic, which gives roughly 34 significant figures. Observe that precision is improved by increasing $\Lambda$ or decreasing $\Delta$, approaching rapidly machine precision when $\Lambda$ is increased. On the other hand, Fig. 3(c) shows the normalized wavefunctions calculated by using the method in §2.2 with energies $E_0$, $E_3$, $E_6$ and $E_9$ (red solid lines), and the relative error (blue dots) when we compare them to the exact ones. Observe the excellent agreement between calculated wavefunctions and their analytical counterparts, with more than 28 significant figures of precision, except if $|x|$ is large, since here $\left|S^{x_0,R}-S_{11}^{x_0,\tilde{x}-\Delta}\right|$ becomes closer to machine precision.

*3.2 Hydrogen atom*

The hydrogen atom has a central potential, and its wavefunctions can be written as

$$\psi_{nlm}(r,\theta,\varphi)=\frac{u_{nl}(r)}{r}Y_l^m(\theta,\varphi). \quad (28)$$

The radial eigenfunction $u_{nl}(r)$ satisfies the 1DSE,

$$-\frac{\hbar^2}{2m}\frac{\partial^2 u_{nl}(r)}{\partial r^2}+V_l(r)u_{nl}(r)=E_{nl}u_{nl}(r) \quad (29)$$

where $V_l(r)$ is the effective potential

$$V_l(r)=\frac{\hbar^2 l(l+1)}{2mr^2}-\frac{e^2}{4\pi\varepsilon_0 r}. \quad (30)$$

Bound states in eq. (28) have boundary conditions $u_{nl}(0) = 0$ and $\lim_{r \to \infty} u_{nl}(r) = 0$, and energies $E_{nl} < 0$. Since there is a regular singularity at $r = 0$, a Frobenius series solution can be obtained, which lead us to analytical energies $E_n^{\text{exact}} = E_1/n^2$, where $n = l+1, l+2, l+3, \cdots$, $E_1 \equiv -\hbar^2/2ma^2$ is the ground state energy, and $a$ is the Bohr radius. However, in other cases, such as the Lennard-Jones potential presented in §4, Frobenius method is not suitable due to an essential singularity at $r = 0$. Hence, we take a different strategy to solve hydrogen atom that can be extended to more general cases. Firstly, note that if $l \geq 1$, the effective potential acts as an infinite barrier near $r = 0$. On the other hand, if $r$ is big enough, the effective potential approximates a constant zero. Then, we may approximate potential (30) as

$$V_l^{aprox}(r) = \begin{cases} \infty & \text{if } r < h_1 \\ \dfrac{\hbar^2 l(l+1)}{2mr^2} - \dfrac{e^2}{4\pi\varepsilon_0 r} & \text{if } h_1 \leq r \leq h_2. \\ 0 & \text{if } h_2 < r \end{cases} \quad (31)$$

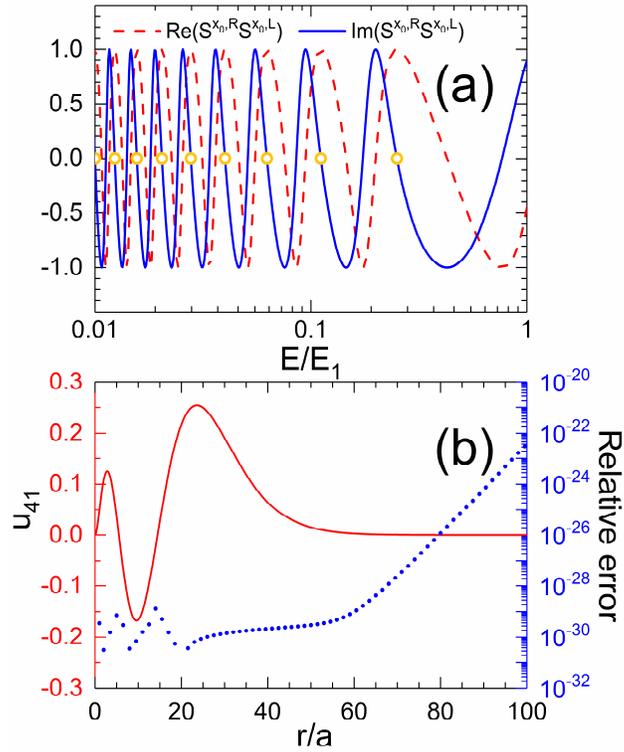

Fig. 4. (a) Calculated imaginary (blue solid line) and real (red dashed line) parts of $S^{x_0,L} S^{x_0,R}$ as functions of energy $E$ for the hydrogen atom with $l = 1$ and $x_0 = 2a$. (b) Calculated radial wavefunction $u_{41}$ (red solid line) and its relative error (blue dots).

Figure 4(a) shows the imaginary (blue solid line) and real (red dashed line) parts of $S^{x_0,L}S^{x_0,R}$ for potential (31) with $l=1$, $h_1 = 9.7844 \times 10^{-11} a$, $h_2 = 20200 a$ by using the method in §2 with $x_0 = 2a$, $V_0 = 1.2 E_1$, $M = 50$ and $\Lambda = M + 2$. Exact energies of Hydrogen atom are represented by yellow open circles. Following the method in Ref. 23, calculations were done by dividing potential in slices centered at points $x_i$ with variable-width $2\Delta_i = 2x_i/A$. In Fig. 3 we use $A = 100$, which ensures that truncated Taylor expansions of potential inside each slice are exact within machine precision. Observe that $S^{x_0,L}S^{x_0,R} = 1$ around the exact energies. In fact, after using the secant root finding method, energies with at least 30 significant figures of precision are obtained for the first nine energies when calculations are done with quadruple precision arithmetic. Figure 4(b) shows the calculated normalized radial wavefunction $u_{41}(r)$ (red line) and the relative error (blue dots) when compared to the analytic exact one, by using the same parameters of Fig. 4(a).

## 4. Lennard-Jones potential

The radial 1DSE for the Lennard-Jones potential, with effective potential

$$V_l(r) = \frac{\hbar^2 l(l+1)}{2mr^2} + 4\varepsilon\left[\left(\frac{\sigma}{r}\right)^{12} - \left(\frac{\sigma}{r}\right)^{6}\right], \quad (32)$$

has an essential singularity at $r = 0$, which disallow us to find analytically its bound states. Observe that for any value of $l$, this potential acts as an infinite potential barrier around $r = 0$ and for large $r$ it approximates a constant zero. Similarly to the hydrogen atom case, we approximate the effective potential (32) as

$$V_l^{aprox}(r) = \begin{cases} \infty & \text{if } r < h_1 \\ \dfrac{\hbar^2 l(l+1)}{2mr^2} + 4\varepsilon\left[\left(\dfrac{\sigma}{r}\right)^{12} - \left(\dfrac{\sigma}{r}\right)^{6}\right] & \text{if } h_1 \leq r \leq h_2. \\ 0 & \text{if } h_2 < r \end{cases} \quad (33)$$

Table 1 shows the lowest nineteen calculated energies of 1DSE with potential (33) obtained by using double and quadruple precision arithmetic, with $l = 0$, $h_1 = 0.22\sigma$, $h_2 = 200\sigma$,

$\varepsilon = 10^4 \hbar^2 / 2^{\frac{4}{3}} m\sigma^2$, $x_0 = 2^{1/6}\sigma$, $V_0 = -1.1\varepsilon$, $M = 50$ and $\Lambda = M + 2$. Potential was divided in slices with center at points $x_i$ and width $2\Delta_i = 2x_i/A$ with $A = 500$. Cases with $n = 4, 6, 8, 10, 12$ and $16$ are in full agreement to those reported by B.D. Shizgal.[5)] Observe that eigenvalues calculated with double precision arithmetic are accurate with at least 14 significant figures when compared to those obtained using quadruple precision arithmetic, showing the numerical stability of this method. Table I also presents the radial expected value $\langle r \rangle$ and standard deviation $\sigma_r$ of the related wavefunction. As expected, wavefunctions with bigger energy have higher $\langle r \rangle$ and more delocalized wavefunctions, due to a greater turning point. This can also be noticed in Figure 5, which displays normalized radial eigenfunctions, $u_{n,0}$, for the cases of $n = 4, 6, 8, 10, 12$ and $16$. In the Supplemental material, we include an implementation of the method to solve the Lennard-Jones potential written in FORTRAN language.[28)]

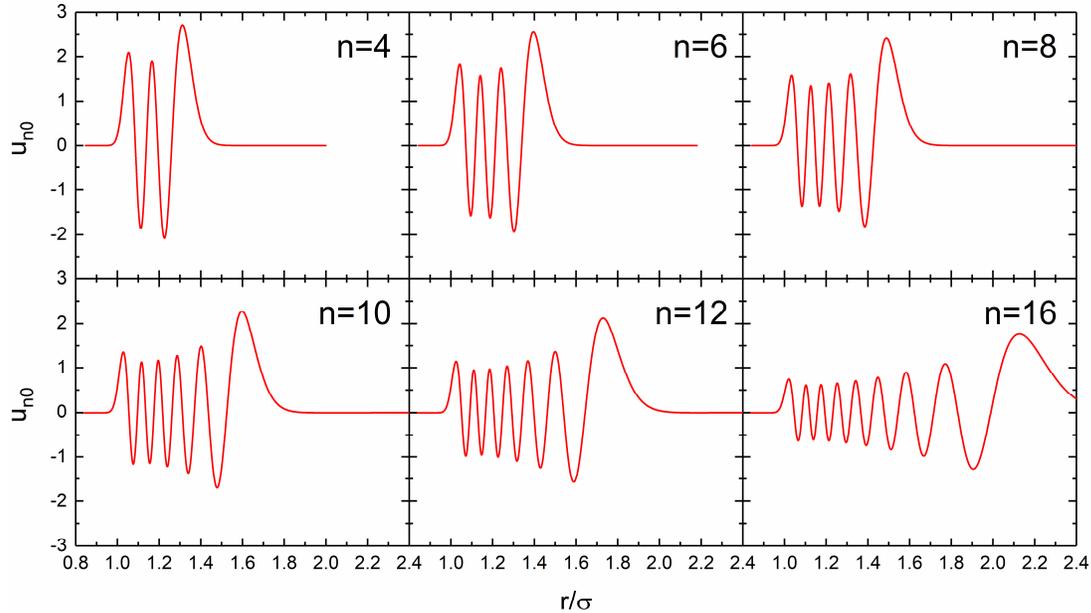

Fig. 5. Calculated $l = 0$ radial eigenfunctions $u_{n,0}$ of Lennard-Jones potential with $\varepsilon = 10^4 \hbar^2 / 2^{\frac{4}{3}} m\sigma^2$.

Table I. Calculated energy ($E_{n0}$), radial expected value ($\langle r \rangle$) and standard deviation ($\sigma_r$) of the first nineteen levels of Lennard-Jones potential with $\varepsilon = 10^4 \hbar^2 / 2^{\frac{4}{3}} m\sigma^2$ and $l = 0$. This potential was approximated by that in eq. (33) with $h_1 = 0.22\sigma$, $h_2 = 200\sigma$. Energies are presented for calculations using Double (DP) and Quadruple (QP) precision arithmetic.

| n | DP $E_{n0}/\varepsilon$ | QP $E_{n0}/\varepsilon$ | $\langle r \rangle / \sigma$ | $\sigma_r / \sigma$ |
|---|---|---|---|---|
| 0 | -0.941046032004322 | -0.941046032004322254854169387438 | 1.13250763 | 0.03332738 |
| 1 | -0.830002082985871 | -0.830002082985871138618852277327 | 1.15362644 | 0.05845972 |
| 2 | -0.727645697519941 | -0.727645697519940580575014256000 | 1.17638781 | 0.07714456 |
| 3 | -0.633692951881524 | -0.633692951881524379652175609420 | 1.20099976 | 0.09349374 |
| 4 | -0.547852043328306 | -0.547852043328305682099735116510 | 1.22770757 | 0.10872536 |
| 5 | -0.469822910169227 | -0.469822910169226681317236593090 | 1.25680268 | 0.12341928 |
| 6 | -0.399296840303147 | -0.399296840303146508761699009100 | 1.2886344 | 0.13792451 |
| 7 | -0.335956071146719 | -0.335956071146718951545689992970 | 1.32362541 | 0.15248893 |
| 8 | -0.279473385016170 | -0.279473385016169764870106170470 | 1.36229270 | 0.16731436 |
| 9 | -0.229511705458584 | -0.229511705458583857183175051710 | 1.40527636 | 0.18258579 |
| 10 | -0.185723701795511 | -0.185723701795510768783095618740 | 1.45337996 | 0.19849100 |
| 11 | -0.147751411297187 | -0.147751411297186707207901024140 | 1.50762849 | 0.21523776 |
| 12 | -0.115225890997835 | -0.115225890997834906020796124040 | 1.56935372 | 0.23307297 |
| 13 | -0.087766914228358 | -0.087766914228358351169241723000 | 1.64032393 | 0.25230831 |
| 14 | -0.064982730496227 | -0.064982730496226650209185743960 | 1.72294802 | 0.27335898 |
| 15 | -0.046469911357580 | -0.046469911357579980075792410100 | 1.82061078 | 0.29680714 |
| 16 | -0.031813309315001 | -0.031813309315000577998098448620 | 1.93825223 | 0.32351356 |
| 17 | -0.020586161355897 | -0.020586161355897098123511894130 | 2.08343401 | 0.35482959 |
| 18 | -0.012350373215634 | -0.012350373215633887547854827020 | 2.26846612 | 0.39303815 |

## 5. Discussion and conclusions

In this work we have presented a novel method to solve eigenpairs of the 1DSE with an arbitrary potential well based on accurate calculations of S-matrices of its subsystems.

The method starts by dividing a general potential well in two independent scattering problems, represented by the left and right associated potentials in Fig. 1. The S-matrices of these potentials are phase changes. We show that there is a bound state energy of the potential well when the product of S-matrices of these associated systems, $S^{x_0,L} S^{x_0,R}$, is one.

The S-matrix of left and right potentials can be determined by dividing them in small slices and obtaining the S-matrix of each slice. As described in Ref. 23, the calculation of the S-matrix of each slice can be performed with arbitrary precision by expressing the potential as a Taylor series centered in the slice. This calculation is done individually for each slice and can be completed in parallel.

To determine the bound state energies, we scan the results of $S^{x_0,L}S^{x_0,R}$ as function of energy, as in Fig. 4(a), to determine where its imaginary part changes of sign and its real part approaches one. Observe that this scanning may be done for a selected range of energies of interest, avoiding wasting of time from computing other energies. It also allows parallelization, since different processors may scan different ranges of energies. Then a regula falsi or a secant method is applied to determine with higher precision the energy for which the imaginary part of this product becomes zero. This process is also parallelizable, because the precision refinement can be performed on each energy individually. This precision refinement can be optimized, by increasing progressively the precision arithmetic. For example, in table 1, double precision results are seeds to calculate quadruple precision ones, taking advantage of the reduced computational time during double precision calculations. If needed, quadruple precision results may also be used as seeds on extended precision calculations.

We also propose a method to find wavefunctions given as piecewise Taylor series, allowing a direct computation of its norm and expected values. Since eq. (21) is susceptible to numerical error, quadruple precision arithmetic is recommended during calculations of the wavefunction.

A comparison of the calculated wavefunctions for quantum harmonic oscillator and hydrogen atom showed in Figs. 3(c) and 4(b) with their analytical counter parts, illustrate the accuracy of this method, with more than 28 significant figures of precision. Results obtained for the Lennard-Jones potential also reflect excellent agreement to those obtained by other state of the art methods.[5]

An implementation of this method to solve eigenstates of Lennard-Jones potential is included in the Supplemental material.[28] Few changes to this code could allow us to find eigenstates of other important potential wells such as Yukawa and Morse/Long-range potentials. The only requirement is to write the potential as piecewise Taylor series, task that can be easily done using automatic differentiation.[25]

Notice that during calculations, all information of the system is contained in $1 \times 1$ or $2 \times 2$ S-matrices, making the method highly efficient in terms of computational memory. On the other hand, this method is not perturbative, allowing us to solve eigenstates in cases where perturbation theory fails. We believe that this method can be extended to two- and three- dimensional systems by using the tight-binding approximation, which is currently under development.

In summary, the proposed method to solve energies and wavefunctions of the one-dimensional Schrödinger equation with an arbitrary potential well is highly efficient, parallelizable, accurate and easy to implement.

**Acknowledgment**

This work has been supported by UNAM-DGAPA-PAPIIT IA106617 and IN116819. Computations were performed at Miztli under project LANCAD-UNAM-DGTIC-329.

*E-mail: carlos@ciencias.unam.mx